%% file: template.tex
\documentclass[cameraready]{Interspeech}

\title{ConformalMOS: Uncertainty-Aware MOS Prediction with Conformal Intervals and Ordinal Modeling}

\author[affiliation={1}, orcid=0000-0001-8185-1314, correspondingauthor]{Kehinde}{Elelu}
\author[affiliation={1}, orcid=0000-0002-5540-7401]{Joshua E.}{Siegel}
\author[affiliation={1}, orcid=0009-0008-0037-0542]{Mohammadali}{Saffary}
\author[affiliation={1}, orcid=0009-0006-3557-8488]{Tashfain}{Ahmed}
\author[affiliation={2}]{Simeon}{Babatunde}
\author[affiliation={2}]{Ebuka}{Okpala}

\address{
    $^1$ Michigan State University, East Lansing, MI, USA \\
    $^2$ Clemson University, Clemson, SC, USA
}

\email{elulukeh@msu.edu, jsiegel@msu.edu, saffarym@msu.edu, ahmedt11@msu.edu, sbabatu@g.clemson.edu, okpalaje@gmail.com}


\usepackage{subcaption}
\usepackage{stfloats}
\usepackage{multirow}

\usepackage{comment}

\begin{document}

\maketitle

\begin{abstract}
    Accurate prediction of human Mean Opinion Scores (MOS) is crucial for evaluating synthetic speech quality in text-to-speech (TTS) and voice conversion (VC). Most MOS predictors output only point estimates and ignore uncertainty, limiting reliability in model selection and deployment. We propose ConformalMOS, a framework that augments MOS prediction with conformal prediction to produce statistically valid prediction intervals with guaranteed coverage under exchangeability, alongside point estimates. During training, MOS labels are modeled in an ordinal-aware manner by converting one-hot targets into Gaussian-smoothed distributions, preserving score ordering. This improves uncertainty calibration, particularly when softmax confidence is overconfident on out-of-distribution speech. On BVCC datasets, ConformalMOS achieves competitive accuracy (MSE = 0.08) while delivering empirically validated coverage, enhancing reliability for real-world TTS and VC deployment.
\end{abstract}

\input{modules/1_introduction}

\input{modules/2_methodology}
\input{modules/3_experimentsetting}

\input{modules/4_result}

\input{modules/5_conclusion}

\section{Acknowledgments}
This work was supported by the Michigan Translational Research and Commercialization (MTRAC) Program, funded by the Michigan Strategic Fund and administered by the Michigan Economic Development Corporation (MEDC). This includes support by Grant Case-48166 of the 21st Century Jobs Trust Fund received through the MSF from the State of Michigan. This report was further supported using Federal funds under award 06-79-06452 (subaward agreement 006) from the U.S. Economic Development Administration, U.S. Department of Commerce. The statements, findings, conclusions, and recommendations are those of the author(s) and do not necessarily reflect the views of the U.S. Economic Development Administration or the U.S. Department of Commerce.

\section{Use of Generative AI Disclosure}
Generative AI tools were used only for limited language editing and writing refinement. All scientific content, methodology, experiments, results, analysis, and conclusions were produced, reviewed, and verified by the authors.

\bibliographystyle{IEEEtran}
\bibliography{mybib}

\end{document}

%% file: modules/1_introduction.tex
\section{Introduction}

Accurately predicting human Mean Opinion Scores (MOS) is essential for evaluating the quality of synthetic speech and music. However, the task remains challenging due to the inherent subjectivity of human ratings and variability introduced by noise conditions, accents, and listener bias \cite{rosenberg2017bias, wells2024experimental}. Large-scale benchmarking studies show that even state-of-the-art systems achieve only moderate utterance-level correlation with human scores \cite{liu2025musiceval, huang2025audiomos}, underscoring the difficulty of reliable automatic MOS estimation.

Recent deep learning approaches leverage transfer learning and self-supervised representations to improve predictive accuracy. For example, UTMOSv2 combines spectrogram-based image encoders with self-supervised speech embeddings \cite{Baba2024}, while waveform–spectrogram fusion and cross-modal encoders further enhance performance \cite{Hoq2025}. Despite these advances, existing systems typically output only point estimates of MOS without quantifying predictive uncertainty. This limitation makes them vulnerable in real-world conditions where unseen acoustic variations degrade reliability. As highlighted by \cite{Wang2024}, the lack of principled uncertainty modeling limits safe deployment in applications such as game development, movie production, and virtual assistants.

Although some probabilistic approaches attempt to model uncertainty \cite{Wang2024, Hoq2025}, they do not provide formal guarantees that predicted intervals contain the true MOS. Given the subjective and noisy nature of human ratings, reliable uncertainty estimation is critical for trustworthy deployment.

To address this limitation, we introduce ConformalMOS, an uncertainty-aware MOS prediction framework built upon existing MOS models \cite{elelucomformalmos, elelu2025convm2d2}. Our method integrates conformal prediction, a distribution-free framework that produces statistically valid prediction intervals with formal coverage guarantees under the assumption of exchangeability between calibration and test data \cite{Shafer2008}. Unlike heuristic uncertainty measures, conformal prediction offers theoretical reliability guarantees.

We further incorporate ordinal-aware modeling to respect the discrete and ranked structure of MOS labels. Instead of treating MOS as purely categorical or continuous, we convert one-hot score labels into Gaussian-smoothed probability distributions over discrete score levels \cite{RitterGutierrez2025}. This captures the partial correctness of neighboring scores and better aligns the training objective with human rating behavior.

The main contributions of this work are:

\begin{itemize}
\item \textbf{Conformal Uncertainty Intervals:} Integration of conformal prediction to produce statistically valid MOS prediction intervals with formal coverage guarantees \cite{Shafer2008}.
\item \textbf{Ordinal-Aware Modeling:} Gaussian-smoothed label distributions that explicitly model MOS ordinality \cite{RitterGutierrez2025}, improving point estimation accuracy.
\item \textbf{Improved Accuracy and Robustness:} On the BVCC dataset, our M2D2 backbone with $\alpha=0.05$ achieves a system-level MSE of 0.08, a 7.0\% reduction from FUSE-MOS \cite{Hoq2025}, while providing calibrated uncertainty for low-confidence and out-of-distribution inputs.
\end{itemize}

%% file: modules/2_methodology.tex
\section{Methodology}

\begin{figure*}[h]
    \centering
    \includegraphics[width=0.8\linewidth]{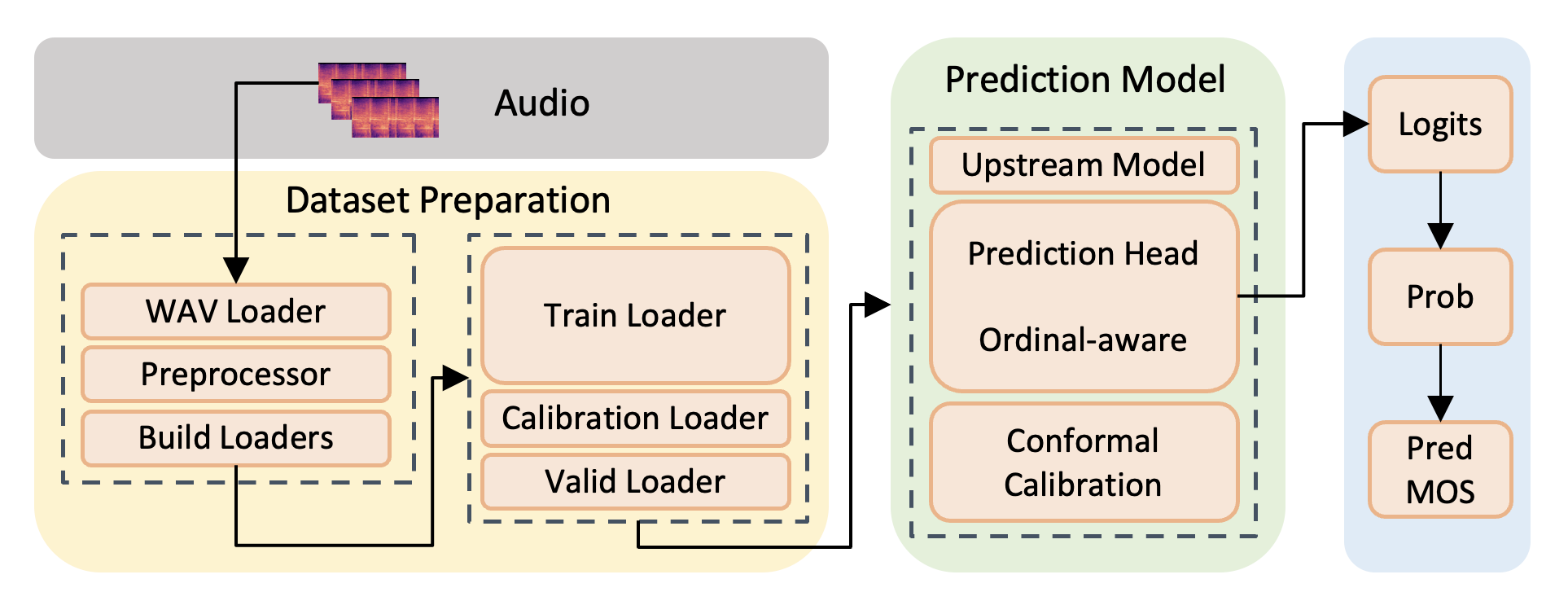}
    \caption{Overview of the proposed ConformalMOS framework. Input audio representations are first processed by a frozen upstream feature extractor. These features are fed into an ordinal prediction head, which uses Gaussian label smoothing to produce point MOS estimates. Finally, post-hoc conformal calibration generates statistically valid prediction intervals, providing reliability-aware predictions.}
    \label{fig:conformalmos_framework}
\end{figure*}

The proposed framework estimates Mean Opinion Scores (MOS) by combining ordinal regression with uncertainty quantification through split conformal calibration. Pretrained audio representations are passed to an ordinal-aware prediction head that outputs a probability distribution over evenly spaced MOS bins. These probabilities are aggregated to produce a scalar MOS estimate. During training, Gaussian-smoothed ordinal targets enforce the ranked structure of MOS labels. In a separate calibration stage, split conformal prediction computes a scalar half-width from calibration residuals, yielding finite-sample valid prediction intervals after clipping to the MOS range (Figure~\ref{fig:conformalmos_framework}).

\subsection{Upstream Model Extraction}

The upstream encoder $\phi$ transforms an input waveform $x$ into a high-level representation suitable for MOS prediction \cite{niizumi2025m2d2}. Rather than learning from raw signals, the downstream head operates on compact embeddings that capture acoustic and prosodic structure. Formally, the encoder produces an embedding $\mathbf{z}\in\mathbb{R}^D$:

\begin{equation}
\mathbf{z} = \phi(x).
\end{equation}

Many pretrained speech models, such as M2D2 or Wav2Vec2, produce frame-level embeddings $\mathbf{Z}\in\mathbb{R}^{T\times D}$, where $T$ denotes the number of frames and $D$ denotes the embedding dimension:
\begin{equation}
\mathbf{Z}=(\mathbf{h}_1,\dots,\mathbf{h}_T).
\end{equation}

We apply temporal mean pooling to obtain a fixed-dimensional representation $\bar{\mathbf{h}}\in\mathbb{R}^D$:
\begin{equation}
\bar{\mathbf{h}}=\frac{1}{T}\sum_{t=1}^{T}\mathbf{h}_t.
\end{equation}

The upstream encoder remains frozen during training to preserve pretrained knowledge and reduce overfitting.

\subsection{Prediction Head}

The prediction head $g_\theta$ maps $\bar{h}$ to (i) logits over $K$ ordinal bins, (ii) class probabilities, and (iii) a scalar MOS estimate \cite{cao2020rank}. We implement $g_\theta$ as a lightweight two-layer MLP with LayerNorm and dropout.

MOS lies in $[1,5]$. Direct regression with $\ell_1$ loss does not align with ranking-based metrics such as SRCC \cite{wang2025qamro}. Instead, we discretize MOS into $K$ equal-width bins spanning $[1,5]$ following \cite{RitterGutierrez2025}, with bin centers $c_k=1+\frac{4(k-1)}{K-1}$ for $k=1,\dots,K$.

This formulation encourages correct relative ordering. To encode ordinality, we replace one-hot labels with Gaussian-smoothed soft targets. For MOS $y\in[1,5]$,
\begin{align}
    \tilde{q}_k(y) =
    \frac{
    \exp\left(-\frac{(y - c_k)^2}{2\sigma^2}\right)
    }{
    \sum_{j=1}^{K}
    \exp\left(-\frac{(y - c_j)^2}{2\sigma^2}\right)
    }.
\end{align}

Following \cite{RitterGutierrez2025}, $\sigma$ is set slightly larger than the bin width to preserve local continuity. This reduces penalties for near-miss predictions while maintaining rank structure across bins.

The scalar prediction is computed as the expected value of bin centers:
\begin{equation}
    \hat{y}(x)=\sum_{k=1}^{K}c_k\,p_k(x),
\end{equation}
ensuring $\hat{y}(x)\in[1,5]$. Training minimizes KL divergence between predicted and smoothed targets, combined with an auxiliary $\ell_1$ loss for stability.

\subsection{Conformal Calibration}

Point estimates alone do not reflect uncertainty. We therefore apply split conformal prediction, which provides finite-sample coverage guarantees without distributional assumptions \cite{lei2013distribution, dhillon2024expected}. The labeled data are divided into a training set and a calibration set (10\%). After training, absolute residuals on the calibration set are computed, and their empirical $(1-\alpha)$ quantile defines the half-width $\hat{q}$. For a new input,
\begin{equation}
    I(x)=\left[\hat{y}(x)-\hat{q},\ \hat{y}(x)+\hat{q}\right]\cap[1,5].
\end{equation}
These intervals satisfy $\mathbb{P}\left(Y\in I(X)\right)\ge 1-\alpha$ under exchangeability. Larger calibration residuals produce wider intervals, signaling higher uncertainty or distribution shift.

Conformal calibration is post-hoc and computationally efficient: residual computation scales linearly with calibration size, and inference requires only adding and subtracting $\hat{q}$ from $\hat{y}(x)$. This provides reliability-aware MOS prediction with minimal overhead.

%% file: modules/3_experimentsetting.tex
\begin{table*}[b]
\centering
\caption{Comparison of FUSE-MOS, prior methods, and Conformal-MOS (Ours) on BVCC datasets. Best values in each column are in \textbf{boldface}.}
\renewcommand{\arraystretch}{1.1}
    \begin{tabular}{l|cccc|cccc}
    \toprule
    \multirow{2}{*}{\textbf{Model}} &
    \multicolumn{4}{c|}{\textbf{Utterance}} &
    \multicolumn{4}{c}{\textbf{System}} \\
    \cline{2-9}
     & MSE↓ & LCC↑ & SRCC↑ & KTAU↑ & MSE↓ & LCC↑ & SRCC↑ & KTAU↑ \\
    \hline
    UTMOS & \textbf{0.165} & \textbf{0.899} & \textbf{0.896} & \textbf{0.730} & 0.090 & 0.936 & 0.936 & 0.795 \\
    FUSE-MOS & 0.191 & 0.887 & 0.887 & - & 0.086 & 0.946 & \textbf{0.947} & - \\
    \hline
    \textbf{Our Models:} \\
    M2D2 ($\alpha=0.01$) & 0.229 & 0.877 & 0.875 & 0.697 & 0.082 & 0.952 & 0.941 & 0.802 \\
    M2D2 ($\alpha=0.05$) & 0.228 & 0.878 & 0.876 & 0.698 & \textbf{0.080} & \textbf{0.953} & 0.943 & \textbf{0.805} \\
    M2D2 ($\alpha=0.1$) & 0.228 & 0.877 & 0.875 & 0.697 & 0.081 & 0.952 & 0.941 & 0.802 \\
    M2D2 ($\alpha=0.2$) & 0.229 & 0.877 & 0.876 & 0.697 & 0.083 & 0.951 & 0.941 & 0.801 \\
    wav2vec ($\alpha=0.01$) & 0.329 & 0.815 & 0.808 & 0.621 & 0.153 & 0.909 & 0.905 & 0.737 \\
    wav2vec ($\alpha=0.05$) & 0.334 & 0.814 & 0.808 & 0.620 & 0.162 & 0.908 & 0.906 & 0.739 \\
    wav2vec ($\alpha=0.1$) & 0.335 & 0.813 & 0.807 & 0.619 & 0.164 & 0.906 & 0.905 & 0.735 \\
    wav2vec ($\alpha=0.2$) & 0.348 & 0.808 & 0.800 & 0.612 & 0.181 & 0.898 & 0.904 & 0.736 \\
    \bottomrule
    \end{tabular}
\label{tab:fuse_vs_conformal_bvcc}
\end{table*}

\section{Experimental Setup}

\subsection{Dataset Preparation}
In our experiments, we utilize the BVCC dataset ~\cite{huang2022voicemos} for training, which comprises 7,106 English audio samples collected from 187 TTS and Voice Conversion (VC) systems included in previous challenges. Although the dataset contains a modest number of samples, it is widely used in speech MOS prediction research and provides diverse samples from multiple TTS and VC systems. Each audio clip is annotated by multiple human raters, ensuring high-quality labels, and such dataset sizes are standard in the field due to the cost and effort of collecting human MOS annotations. For model training and evaluation, we split the BVCC dataset according to the standard partitions defined in the VoiceMOS Challenge: 1,066 samples are reserved for testing, and 1,066 samples for validation. Of the remaining 4,974 samples in the development pool, we reserved a calibration subset of 497 samples (10\%) for conformal interval estimation, leaving 4,477 samples for the final model training. 

\subsection{Training Details}
MOS prediction models were trained on BVCC using NVIDIA RTX A5000 GPUs (24 GB) for up to 1000 epochs with early stopping (patience 20) monitored on validation loss; most runs stopped after 300–400 epochs. Batch size was 32 for training and 8 for validation/testing \cite{masters2018revisiting}. We used SGD with momentum (LR $1\times10^{-4}$, momentum 0.9, weight decay $1\times10^{-4}$) and froze the upstream encoder, updating only the prediction head. Checkpoints were saved at lowest validation loss. Training typically took ~2 hours per run, balancing stable convergence and generalization while preventing overfitting.

\subsection{Evaluation Metrics}
To assess point-prediction quality, we report the following standard evaluation metrics for MOS at both the utterance and system levels:
\begin{itemize}
    \item \textbf{Mean Squared Error (MSE):} Measures the average squared difference between predicted and ground-truth MOS scores.
    \item \textbf{Linear Correlation Coefficient (LCC):} Quantifies the linear relationship between predictions and true scores.
    \item \textbf{Spearman's Rank Correlation Coefficient (SRCC):} Evaluates the monotonic ordering of predictions with respect to ground-truth MOS.
    \item \textbf{Kendall Tau Rank Correlation Coefficient (KTAU):} Measures the ordinal association between predicted and true rankings.
\end{itemize}

To assess the quality of our uncertainty estimates, we evaluate the conformal prediction intervals using the following metrics:

\begin{itemize}
    \item \textbf{Coverage:} Measures the proportion of test samples for which the ground-truth MOS falls within the predicted interval $I(x)$, indicating how well the intervals capture the true values \cite{chen2021learning, kompa2021empirical}:
    \begin{equation}
        \text{Coverage} = \frac{1}{N} \sum_{i=1}^{N} \mathbf{1}\{ y_i \in I(x_i) \},
    \end{equation}
    where $N$ is the number of test samples. Coverage is compared to the nominal level $1-\alpha$.

    \item \textbf{Calibration Error:} Measures the absolute deviation between empirical coverage and the nominal coverage, reflecting how accurately the intervals are calibrated \cite{stephen2021gentle, kompa2021empirical}.
    \begin{equation}
        \text{Calibration Error} = \big| \text{Coverage} - (1-\alpha) \big|
    \end{equation}

    \item \textbf{Average Interval Width:} Represents the mean width of the conformal intervals across all samples, capturing the overall sharpness of the uncertainty estimates \cite{chen2021learning, jonkers2026reliable}.
    \begin{equation}
        \text{Average Width} = \frac{1}{N} \sum_{i=1}^{N} \big( \overline{y}_i - \underline{y}_i \big),
    \end{equation}
    where $I(x_i) = [\underline{y}_i, \overline{y}_i]$.

    \item \textbf{Sharpness (RMS Radius):} Measures a model’s overall uncertainty level on a dataset through the root-mean-square of the half-widths of the intervals, providing an alternative concentration measure \cite{jonkers2026reliable}.
    \begin{equation}
        \text{Sharpness (RMS)} = \sqrt{ \frac{1}{N} \sum_{i=1}^{N} \left( \frac{\overline{y}_i - \underline{y}_i}{2} \right)^2 }
    \end{equation}
\end{itemize}

%% file: modules/4_result.tex
\begin{figure*}[t]
    \centering
    \begin{subfigure}[b]{0.48\linewidth}
        \centering
        \includegraphics[width=\linewidth]{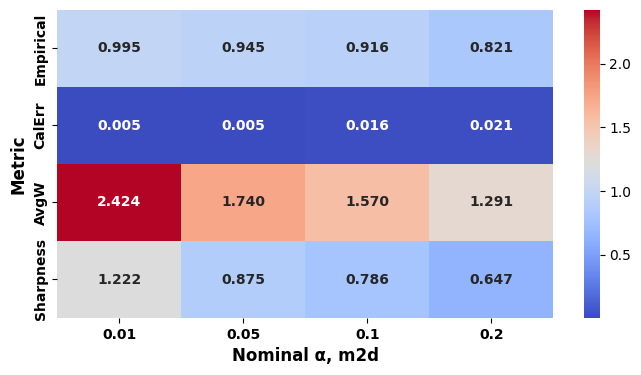}
        \caption{M2D2 Result}
        \label{fig:m2d_result}
    \end{subfigure}
    \hfill
    \begin{subfigure}[b]{0.48\linewidth}
        \centering
        \includegraphics[width=\linewidth]{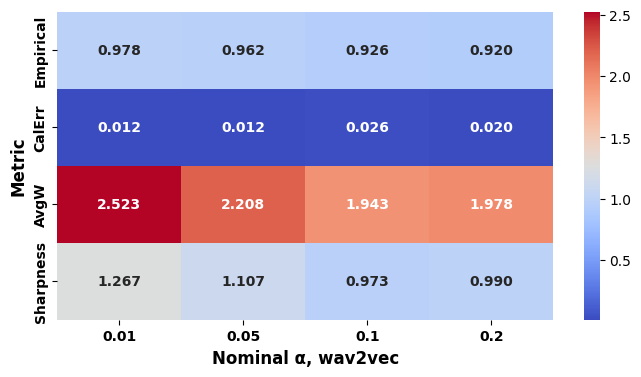}
        \caption{Wav2Vec Result}
        \label{fig:wav2vec_result}
    \end{subfigure}
    \caption{Calibration performance of Conformal-MOS on the BVCC dataset using two different backbones. (a) M2D2 achieves empirical coverage closely matching nominal targets across multiple $\alpha$ levels, with low calibration error and relatively narrow (sharp) intervals, indicating reliable uncertainty estimates. (b) Wav2Vec shows larger calibration error, empirical coverage that falls short at low $\alpha$ levels, and wider intervals, highlighting less reliable uncertainty estimation.}
    \label{fig:side_by_side_results}
\end{figure*}

\section{Experimental Results}

\subsection{Evaluating ConformalMOS on BVCC}

We assess the performance of our proposed ConformalMOS framework on the BVCC dataset at both the utterance and system levels, and compare it with prior state-of-the-art models (UTMOS and FUSE-MOS). Table~\ref{tab:fuse_vs_conformal_bvcc} summarizes the results. We first observe that UTMOS and FUSE-MOS provide strong baselines, achieving low MSE and high correlation metrics. While UTMOS attains slightly better utterance-level accuracy, FUSE-MOS performs best in system-level ranking. By incorporating the upstream model M2D2, conformal prediction, and ordinal-aware modeling into the MOS estimation pipeline, our ConformalMOS models achieve competitive or superior system-level performance. In particular, the ConformalMOS M2D2 models consistently reach the highest system-level LCC while maintaining the lowest MSE. Our evaluation is limited to in-domain BVCC data, so out-of-domain robustness remains untested; we also omit heuristic comparisons because they lack conformal
prediction’s finite-sample coverage guarantee.

We evaluated multiple $\alpha$ values 0.01, 0.05, 0.1, and 0.2 to examine their impact on conformal prediction intervals for MOS estimation. The choice of $\alpha$ directly affects the width of the intervals and the associated coverage probability. Larger $\alpha$ values produce narrower intervals (improved sharpness) but increase the risk of undercoverage, while smaller $\alpha$ values yield wider intervals to ensure higher nominal coverage. Selecting multiple levels allows us to study the trade-off between interval tightness and reliability, aiming for intervals that are both informative and statistically valid. Empirical studies have shown that appropriate selection of $\alpha$ is essential for controlling error rates and ensuring valid coverage in conformal prediction \cite{chernozhukov2018exact}. Based on these considerations, we selected an $\alpha$ of 0.05 as our primary setting, corresponding to 95\% nominal coverage, which balances interval width with reliability while maintaining the best overall performance. This shows that the novel pipeline improves the reliability of predictions without degrading point-estimation accuracy. At the utterance level, ConformalMOS M2D2 maintains strong correlations close to UTMOS and FUSE-MOS, with only a slight increase in MSE. In contrast, the ConformalMOS wav2vec models underperform notably across all metrics, indicating that the choice of backbone strongly affects performance under the conformal framework. Overall, these results highlight that conformal and ordinal prediction enhance system-level robustness, especially when combined with the M2D2 backbone. Our best-performing ConformalMOS M2D2 $\alpha = 0.05$ configuration surpasses baselines at the system level, while retaining competitive utterance-level accuracy. This confirms that conformal prediction can be effectively integrated into MOS estimation pipelines to improve prediction reliability without compromising prediction quality.

\subsection{Backbone Effects on Conformal-MOS Calibration}
Figures \ref{fig:m2d_result} and \ref{fig:wav2vec_result} present the calibration performance of our Conformal-MOS models on the BVCC dataset. Ideally, empirical coverage should closely match the nominal coverage, calibration error should be small, and intervals should be as narrow (sharp) as possible while remaining well-calibrated. The M2D2 models exhibit consistently accurate coverage. Empirical coverage aligns closely with nominal targets at most $\alpha$ levels, with low calibration error, especially for the M2D2 backbone. As expected, increasing $\alpha$ leads to narrower prediction intervals and improved sharpness. This demonstrates the expected coverage-sharpness trade-off and confirms that the conformal calibration procedure is functioning properly on this backbone. In contrast, the wav2vec models show higher calibration error and less precise coverage, particularly at lower $\alpha$ values. During the $\alpha$ of $0.01$, the empirical coverage falls short of the target 0.99, and the intervals are much wider with higher sharpness values. This indicates that the weaker wav2vec backbone produces less reliable uncertainty estimates under the conformal framework. Overall, these results confirm that conformal prediction produces well-calibrated uncertainty intervals when combined with a strong backbone like M2D2, while performance degrades when the underlying point predictions are less accurate. This highlights the importance of backbone choice for reliable uncertainty estimation.

%% file: modules/5_conclusion.tex
\section{Conclusion}

We introduce ConformalMOS, a reliability-aware MOS prediction framework combining ordinal modeling with split conformal calibration. Unlike standard regression producing only point estimates, ConformalMOS outputs prediction intervals with formal coverage guarantees, capturing the subjectivity and variability of human MOS ratings. Experiments on BVCC show improved system-level performance with competitive utterance-level accuracy. The M2D2 backbone yields well-calibrated intervals with empirical coverage matching nominal targets, while wav2vec shows wider intervals and higher calibration error, emphasizing the need for strong base predictors. Analysis confirms the coverage–sharpness trade-off: increasing $\alpha$ narrows intervals but reduces coverage. By integrating ordinal-aware training with conformal uncertainty, ConformalMOS enhances interpretability and reliability, offering a practical, theoretically grounded approach for trustworthy MOS prediction in real-world applications. Listener disagreement and rater-specific effects remain important future work.